\documentclass[fleqn,usenatbib]{mnras}   

\usepackage[final]{changes}
\usepackage{graphicx}
\usepackage{amsmath}
\usepackage{amssymb}
\usepackage{booktabs}
\usepackage{latexsym}
\usepackage{multirow}
\usepackage{hyperref}

\hypersetup{
    colorlinks,
    citecolor=blue,
    filecolor=blue,
    linkcolor=blue,
    urlcolor=blue
}
\colorlet{Changes@Color}{orange}

\title[Stellar Influence on Ion Escape]{Stellar Influence on Heavy Ion Escape from Unmagnetized Exoplanets}

\author[H. Egan]{Hilary Egan$^{1}$\thanks{Contact e-mail: \href{mailto:Hilary.Egan@colorado.edu}{Hilary.Egan@colorado.edu}},
 Riku Jarvinen$^{2,3}$, David Brain$^{1}$
\
\\
$^{1}$ Department of Astrophysical and Planetary Sciences, University of
   Colorado, Boulder, CO 80309, USA ,\\
$^{2}$ Department of Electronics and Nanoengineering, School of Electrical Engineering, Aalto University, Espoo, Finland,\\
$^{3}$ Finnish Meteorological Institute, Helsinki, Finland}

\date{Revised 2019 March 11; submitted 2019 January 5}

\pubyear{2019}

\begin{document}
\label{firstpage}
\pagerange{\pageref{firstpage}--\pageref{lastpage}}
\maketitle

\begin{abstract}

Planetary habitability is in part determined by the atmospheric evolution of a planet; one key component of such evolution is escape of heavy ions to space. Ion loss processes are sensitive to the plasma environment of the planet, dictated by the stellar wind and stellar radiation. These conditions are likely to vary from what we observe in our own solar system when considering a planet in the habitable zone around an M-dwarf. Here we use a hybrid global plasma model to perform a systematic study of the changing plasma environment and ion escape as a function of stellar input conditions, which are designed to mimic those of potentially habitable planets orbiting M-dwarfs. We begin with a nominal case of a solar wind experienced at Mars today, and incrementally modify the interplanetary magnetic field orientation and strength, dynamic pressure, and Extreme Ultraviolet input. We find that both ion loss morphology and overall rates vary significantly, and in cases where the stellar wind pressure was increased, the ion loss began to be diffusion or production limited with roughly half of all produced ions being lost. This limit implies that extreme care must be taken when extrapolating loss processes observed in the solar system to extreme environments.

\end{abstract}

\section{Introduction}

Recent developments in exoplanet observation techniques have allowed the discovery of thousands of extra-solar planets, including dozens of small, rocky planets that are potentially habitable. The closest star to us, Proxima Centauri, hosts a planet with a minimum mass of 1.3 $M_E$ \citep{2016Natur.536..437A}, and the nearby Trappist system is home to seven transiting Earth sized planets, three or four of which are in the habitable zone (HZ) of the star \citep{2017Natur.542..456G, 2016Natur.533..221G}. Analysis of the Kepler data has shown that planetary systems are common around M-dwarfs \citep{2013ApJ...767L...8K}, and these systems also show the best promise of observing exoplanet atmospheres \citep{2016PhR...663....1S}.

As planetary atmospheres affect the surface temperature and prevent rapid water loss, atmospheric evolution of terrestrial planets around M-dwarfs is a topic of key importance. Atmospheric evolution can encompass a broad variety of processes, including volcanic out-gassing, sequestration, and loss to space. One component of loss to space is thermal loss, where particles with a thermal energy exceeding the escape velocity of the planet escape; however, heavier elements with higher escape velocities will have more difficult escaping thermally. Non-thermal processes, including those that act on ions, act to increase the energy available to a given particle and therefor provide additional paths to escape for heavy ions.

Non-thermal loss processes have been studied extensively for solar system planets including Earth \citep[e.g][]{2005JGRA..110.3221S}, Mars \citep[e.g][]{1989Natur.341..609L, 2007Sci...315..501B, 2015GeoRL..42.9142B}, Venus \citep[e.g][]{2013JGRA..118.3592N, 2007Natur.450..650B}, and Titan \citep[][e.g]{2005Sci...308..986W, 1982JGR....87.1395G}. This loss takes a variety of observed forms including photo-chemical escape \citep{1994Icar..111..271J,2009Icar..204..527F}, charge exchange \citep{1977JGR....82....1C}, sputtering \citep{1994Icar..111..271J, 1993P&SS...41..657L,2001P&SS...49..645L}, ion pickup \citep{1991JGR....96.5457L}, ion bulk escape \citep{2010GeoRL..3714108B}, and the polar wind \citep{1968P&SS...16.1019B, 2007JASTP..69.1936Y}.

Further understanding of ion loss from terrestrial solar system planets has been developed using 3D global plasma models. These models are useful as they allow one to probe the state of the whole system and its drivers at once, rather than limited in situ observations from spacecraft. Plasma models such as magnetohydrodynamic (MHD) \citep[e.g][]{1998JGR...103.4723K, 2002JGRA..107.1282M, 2013JGRA..118..321M, 2009JGRA..114.9208T}, hybrid \citep[e.g][]{1991JGR....9611209B, 2002JGRA..107.1035K, 2002JGRA..107.1471T, 2005AnGeo..23..433M, 2009P&SS...57.2001S,2009AnGeo..27.4333J}, and test particle/direct simulation Monte-Carlo (DSMC) methods \citep[e.g][]{2002JGRA..107.1170C,2008JGRA..113.2210F,Luhmann2006}, have all been used to understand ion escape in the context of the terrestrial planets. 

Due to the relative abundance of planetary systems and constraints from the signal to noise ratio in most observing techniques, the most potentially observable planets that meet this criteria are found in the habitable zone around M-dwarfs. These environments are likely to be extreme due to the enhanced EUV \citep{2016ApJ...820...89F} and the closer radius of the habitable zone relative to solar. Each of these factors is likely to have a distinct effect on the ion loss of the planet, and it is necessary to understand how they work in conjunction.

Using plasma simulations that have been validated in the solar system planetary context can add to understanding of ion loss in exoplanetary systems as well as young solar system planets \citep{2011A&A...525A.117J,doi:10.1089/ast.2008.0250,2010P&SS...58.2031B}. Exoplanets may differ from solar system planets in their intrinsic properties such as size, composition, or magnetic field, as well as the external conditions dictated by the interaction with the host star. \added{\citet{2015ApJ...806...41C} explore the influence of an M-dwarf star on a Venus-like planet in the habitable zone, concentrating on the impact the possible sub- and super-alfvenic stellar wind. \citet{2017ApJ...844L..13G} examine the influence of a magnetic field in the protection of a planet from atmospheric escape in the habitable zone of red dwarf Proxima Centauri using a 1D polar wind outflow model.}

\replaced{Here we explore the case of an unmagnetized planet orbiting in the habitable zone of a generalized M-dwarf system.}{Planets in the habitable zone around M-dwarfs are likely to be unmagnetized or weakly magnetized \citep{2010ApJ...718..596G}}. Although magnetospheres are classically considered necessary to prevent solar wind erosion of atmospheres, this may not necessarily be the case. Estimates of ion escape from Mars, Venus, and Earth all show similar rates \citep{2005JGRA..110.3221S,2013cctp.book..487B}, despite Earth's strong intrinsic magnetic field and the lack thereof at Mars and Venus. Thus it is still worth considering and very necessary to study the plasma environment and ion escape of unmagnetized planets.

Here we present a systematic study of the changing plasma environment and planetary ion escape as a function of stellar input conditions. The input conditions have been selected to incrementally change from typical solar wind today to the stellar wind at potentially habitable planets orbiting M-dwarfs. We begin with a base case of Mars today, and alter the interplanetary magnetic field (IMF) orientation, dynamic and magnetic pressure, and EUV flux. Section \ref{sec:stellar_params} describes the choices in stellar input conditions, Section \ref{sec:methods} describes the model, Section \ref{sec:results} gives our results, Section \ref{sec:discussion} further discusses the limitations and implications of our results, and Section \ref{sec:conclusions} summarizes our conclusions.

\section{Stellar Parameters}
\label{sec:stellar_params}

Both the intrinsic stellar parameters and the habitable zone location drive differences in the stellar influence on terrestrial planet escaping atmospheres. Here we describe some of the general differences and motivate our selection of parameters. Our initial base case (R0), is the same as that studied by \citet{doi:10.1029/2017JA025068}, and is an example of a typical solar wind experienced by Mars. The final case (R4), is identical to the case considered for Trappist-g by \citet{2018PNAS..115..260D}, where the stellar wind was reconstructed using the Alfven Wave Solar Model \citep{2014ApJ...782...81V}.

\subsection{Quasi-Parallel IMF}

For unmagnetized planets, much of the interaction with the solar wind is dictated by the direction of the IMF. Because ions are constrained to gyrate around magnetic fields, both solar particle inflow and planetary ion outflow will change due to the influence of the IMF. An interaction with the solar wind and unmagnetized planets (eg. Venus and Mars) is typically sketched with magnetic field lines roughly perpendicular to the direction of solar wind flow piling up around the induced ionosphere and then slipping past the planet. However, configurations where the magnetic field is more aligned than perpendicular with the solar wind flow occur in the inner solar system due to the Parker spiral structure of the IMF and occur at exoplanets orbiting close to their host stars. Aligning the magnetic field with the solar wind flow will create regions where ions can flow directly away from the planet along field lines normal to the planet surface, dramatically changing the ionospheric interaction \citep{2011A&A...525A.117J, 2009AdSpR..43.1436L, doi:10.1029/2009GL040515}. Furthermore, the radial magnetic field results in vanishing upstream convection electric field, which is the large-scale energy source for ion pickup \citep[e.g.][]{2014JGRE..119..219J}.

Additionally, a shock is unstable when the angle between the magnetic field and the local shock normal $<15^\circ$ \citep{2008arXiv0805.2579T, 2008arXiv0805.2162T}. A quasi-parallel magnetic field will satisfy this condition for an entire hemisphere of the bow shock, thus making the interaction quite different than the quasi-perpendicular IMF. \replaced{Because quasi-parallel shocks do not form stable well-defined surfaces and can reflect ions in an extended foreshock region, kinetic effects associated with finite ion gyroradii and the ion velocity distribution become important \citep{2008arXiv0805.2579T, 2008arXiv0805.2162T}. This makes hybrid models well-suited to simulating such an interaction.}{This makes the kinetic effects related to ion motion important, indicating that a hybrid model is better suited to simulating such an interaction than an MHD model }.

As M-dwarfs are relatively dimmer than the Sun, the habitable zone must be correspondingly closer in. In addition to causing increased stellar fluxes, this will also likely lead to more radially oriented IMFs as expected from a Parker spiral model. MHD simulations of M-dwarf stellar winds such as \citet{2017ApJ...843L..33G} also show radially oriented IMFs in the corresponding habitable zones. Although it is not universally true that all exoplanets in the HZ of an M-dwarf will experience a radial IMF, it is a significant departure from what the potentially habitable solar system planets (Venus, Earth, and Mars) experience so it is helpful to investigate its effects. 

Here we consider a quasi-parallel magnetic field, with $\alpha=18.2$ (R1), which in the context of the solar system is similar to the nominal Parker spiral angle at Mercury. Considering a perfectly radial field is both somewhat unlikely to the slight variability in both the solar wind and the IMF, and more difficult computationally due to instabilities in the ionospheric interaction and vanishing upstream convection electric field. \added{Additionally, although the Parker spiral angle for the Trappist-1 exoplanets would be far less than the chosen value, this value is motivated by the solar wind reconstruction model used by \citet{2018PNAS..115..260D}.

This choice does, however, neglect the influence of the large orbital velocities these planets must have, which will be comparable to the stellar wind velocity. Thus the corresponding angle of the stellar wind as seen by the planet and the magnetic field will also change. The results we present here neglect this effect for ease in comparison of results, however, future work could study the influence of the planetary orbital velocities.}

\subsection{Solar Wind Strength}

Solar wind momentum is key source of energy input into the planetary plasma environment. Observations of terrestrial planets have shown that ion loss is dependent on solar wind dynamic pressure \citep{doi:10.1002/2017JA024306,2017EGUGA..1911139B}.

It is not currently possible to measure the stellar wind of stars besides our Sun directly, so investigations into these effects rely on stellar wind reconstructions using MHD solar wind models \citep[e.g.][]{2015MNRAS.449.4117V,2017ApJ...843L..33G}. 

The steady state stellar wind may vary extensively across a single orbit for a close in planet \citep{2017ApJ...843L..33G, 2016ApJ...833L...4G}. Furthermore, space weather may also increase variability. Previous models of exoplanet ion loss have investigated the steady state loss rates for two cases: maximum total pressure, and minimum total pressure \citep[e.g.][]{2018PNAS..115..260D}, effectively varying the magnetosonic mach number. 

Here we consider the scaling of stellar wind in two parts, increasing the overall pressure, and varying the ratio of magnetic to dynamic pressure. We first scale the overall pressure by a factor of roughly $4\times10^3$ (R2), and then increase the solar wind density by a factor of 100 (R3), decreasing the ratio of magnetic to dynamic pressure. \replaced{This mimics an overall increase due to the increased flux expected for a closer habitable zone distance and then a possibly extreme dynamic pressure dominated scenario. While an actual planet ma33y experience extreme variation in stellar wind due to both orbital variation and the intrinsic variability of the wind, here we select two interesting cases for study.}{Although our aim is to isolate the effect of each parameter, it is not possible to treat magnetic and dynamic pressure completely independently due to numerical stability constraints.}

\subsection{EUV Input}

Another critical component of the stellar interaction with a planetary atmosphere is the input in the UV and Extreme UV (EUV). In addition to photoionization of planetary neutrals, EUV photons are absorbed in the upper atmosphere leading to heating, and in some cases thermal driven hydrodynamic escape \citep{1987Icar...69..532H}. In cases where heavy elements like Oxygen are gravitationally bound to the atmosphere, EUV input is still correlated with ion escape rates \citep{doi:10.1002/2017JA024306}.

\replaced{EUV flux will also increase due to the proximity of the habitable zone, similarly to the stellar wind flux. Furthermore, observations show that M-dwarfs have EUV fluxes of 10-1000 times that of solar \citep{2016ApJ...820...89F}. Here we chose to scale  $I_{EUV}$ by a factor of 100 (R4).} {Observations show that M-dwarfs have EUV fluxes of 10-1000 times that of solar \citep{2016ApJ...820...89F}, so here we chose to scale $I_{EUV}$ by a factor of 100 (R4).} Although we do not directly simulate the stellar radiation environment, our overall ion production rate scales directly with $I_{EUV}$. For further description of this implementation see Section \ref{sec:methods}. \added{While stellar activity may dominate the EUV flux experienced by the planet, we here examine a steady state case.}

\subsection{Summary}

Table \ref{tab:stellar_params} summarizes the stellar parameters used for our suite of simulations. Each simulation builds upon the changes of the last, such that R2 contains the same adjustments as R1, R3 contains the R2 and R1 adjustments and so on. We also list a variety of relevant plasma scales that further illustrate the differences and similarities between the models.

\begin{table*}

\centering
\begin{tabular}{@{}llllll@{}}
\toprule
Simulation                                                       & R0 : Nominal                      & R1 : Parallel-IMF                          & R2 : Total-Pressure                          & R3 : Density                 & R4 : EUV                 \\ \midrule
$u$ (km/s)                                                       & 350                      & 350                               & \textbf{604}                 & 604                 & 604                 \\
$T$ (K)                                                          & 5.91e4                   & 5.91e4                            & \textbf{1.26e6}              & 1.26e6              & 1.26e6              \\
$n(H^+)$ (cm$^{-3}$)                                               & 4.85                     & 4.85                              & \textbf{6.44e2}              & \textbf{5.79e3}     & 5.79e3              \\
$B$ (nT)                                                         & {[}-0.74, 5.46, -0.97{]} & \textbf{{[}-5.31, 0.44, -1.51{]}} & \textbf{[-149, 13, -42]} & {[}-149, 13, -42{]} & {[}-149, 13, -42{]} \\
$Q({O}^+)$    ($10^{25}/s$)       & 2                        & 2                                 & 2                            & 2                   & \textbf{200}        \\
$Q({O_2}^+)$  ($10^{25}/s$) & 1.4                      & 1.4                               &  1.4                            & 1.4                 & \textbf{140}        \\
\midrule

$|B|$ (nT)                                                       & 5.59                     & 5.59                              & \textbf{155}                 & 155                 & 155                 \\
$\alpha$ ($^\circ$)                                              & 82.4                     & \textbf{18.2}                     & 18.2                         & 18.2                & 18.2                \\
$v_A$ (km/s)                                                     & 55.3                     & 55.3                             & \textbf{133}                          & \textbf{44.4}                & 44.4                \\
$M_A$                                                            &          6.3                &       6.3                             & \textbf{4.5}                              &     \textbf{13.6}                &       13.6              \\
$P_B/P_{dyn}$                                                          &        0.02                  & 0.02                                   &  0.02                            &   \textbf{0.005}                  &   0.005                  \\
$r(O^+)$ (km)                                     & 10364                    & \textbf{3266}                              & \textbf{203}                          & 203                 & 203                 \\
$r({O_2}^+)$ (km)                              & 20728                    & \textbf{6532}                              & \textbf{406}                          & 406                 & 406                 \\
$\bar{E}(O^+)$ (keV) & 19.5 &1.6&4.9&4.9&4.9 \\
$\bar{E}({O_2}^+)$ (keV) & 38.9 &3.2 &9.8&9.8&9.8\\
 \bottomrule
\end{tabular}
\caption{Parameters used to drive each simulation. Simulations are labelled R0 through R4. Parameters that are directly configured in the simulation are listed on top, while derived parameters are listed below. Numbers in \textbf{bold} are changed from the preceding simulation.}
\label{tab:stellar_params}
\end{table*}

\emph{Stellar wind speed} ($u$): Input speed of the incident stellar wind in the $-x$ direction.

\emph{Temperature} ($T$): Temperature of the incident solar wind Hydrogen ions.

\emph{Number Density} ($n(H_+)$): number density of the incident solar wind Hydrogen ions.

\emph{IMF} ($B$): Incident stellar wind magnetic field vector, in PSE coordinates (described in Section \ref{sec:methods}).

\emph{Production Rate} ($Q$): Total production rate for a given ion (described further in Section \ref{sec:methods}).

\emph{IMF Angle} ($\alpha$): angle between the upstream stellar wind velocity and the IMF, smaller angle indicates a more parallel interaction.

\emph{Alfven Speed} ($v_A = B/\sqrt{\mu_0\rho}$, $\mu_0 :=$ magnetic permitivity, $\rho :=$ density): Alfven speed in the incident solar wind

\emph{Alfven Mach Number} ($M_A = u/v_A$): Determines the nature of the bow shock.

\emph{Magnetic to dynamic pressure ratio} ($P_B/P_{dyn}=(B^2/2\mu_0)/(1/2\rho u^2)$): Ratio of the incident solar wind magnetic pressure and dynamic pressure, influences how the magnetic field lines drape around the planet.

\emph{Ion Gyroradius} ($r=mv_\perp/qB$, $v_\perp:=$ velocity component perpendicular to the magnetic field): dictates the radius at which an ion moving with the velocity of the upstream solar wind gyrates in the upstream magnetic field. This has an influence on the trajectory of escaping particles when the radius is comparable to the size of the planet (3390 km). 

\emph{Ideal Gyro-Averaged Pickup Energy} ($\bar{E}=2(1/2 m_i v_\perp)$) ideal energy of an ion gyrating in the solar wind averaged over a gyroperiod, see discussion in \citet{2014JGRE..119..219J}.

\section{Methods}
\label{sec:methods}

The following simulations were performed using RHybrid \citep{RHybrid}, a hybrid global plasma model for planetary magnetospheres. In a hybrid model the ions are treated as macroparticle clouds that are evolved according to the Lorentz equation, while the electrons are treated as a charge neutralizing fluid. This allows the simulation to include ion kinetic effects which are important in situations where the ion gyroradius is large compared to the scale size of the system.

Each ion macroparticle represents a group of ions that have the same velocity ($v_i$), central position ($x_i$), charge ($q_i$), and mass ($m_i$) obeying the Lorentz force such that
\begin{equation}
m_i \frac{d\vec{v_i}}{dt} = q_i(\vec{E}+\vec{v_i}\times\vec{B})\;,
\end{equation}

where $\vec{E}$ and $\vec{B}$ are the electric and magnetic fields respectively. The electron charge density then follows from the quasi-neutral assumption when summed over all ion species.

 The current density is calculated from the magnetic field via Ampere's Law 
 \begin{equation}
\vec{J} = \nabla\times\vec{B}/\mu_0 \;,
\end{equation} and the electric field can then be found using Ohm's law 
	 
\begin{equation}
\vec{E} = -\vec{U_e}\times \vec{B}+\eta\vec{J}\;,
\end{equation}, where the $\eta$ is explicit resistivity profile used to add diffusion in the propagation of the magnetic field \citep{2008SSRv..139..143L}. \added{This value was chosen to be $\eta = 0.02\mu_0\Delta x^2/\Delta t$, such that the magnetic diffusion time scale $\tau_D = \mu_0 L_B^2/\eta=50\Delta t$, for the magnetic length scale $L_B\approxeq\Delta x$. This is a similar value as used in earlier work \citep{doi:10.1029/2017JA025068, RHybrid}, and ensures that the magnetic field diffuses on timescales longer than the timestep $\Delta t$. The explicit resistivity allows some diffusion to stabilize the numerical integration and is greater than the inherent numerical resistivity of the code for the chosen resolution. At the same time $\eta$ is kept small enough to keep the solution from becoming smoothed out by diffusion.} Note that the resistivity is not explicitly included in the Lorentz force. 

Finally, the magnetic field is then propagated using Faraday's Law 
\begin{equation}
\frac{\partial \vec{B}}{\partial t} = -\nabla \times \vec{E}\;.
\end{equation}

See \citet{RHybrid} and references therein for details of the numerical scheme. 

The lower boundary is located at 250 km altitude and is implemented as a super-conducting sphere. This mimics the effect of the electromagnetic properties of the induced magnetosphere. 

In the RHybrid runs analyzed in this work the emission of ionospheric ions in the induced magnetosphere is implemented using a Chapman profile which arises naturally when considering an isothermal atmosphere that is ionized by plane-parallel, monochromatic radiation in the EUV \citep{1957SCoA....2....1C}. The production rate of ions is given by 
\begin{equation}
	q(\chi, z') = Q_0 \exp[1-z'-\sec(\chi)*e^{-z'}]\;,
\end{equation}

where $z'$ is the normalized height parameter given by $z' = (z-z_0)/H$ where $\chi$ is the solar zenith angle, $z_0$ is the reference height, and $H$ is the scale height. In each simulation we use a reference height of 300 km and a scale height of 16 km. We also add an additional constant ionization source behind the planet that is continuous across the terminator to mimic other ionization sources and prevent divergence caused by extremely low densities.

We note that this is not a self-consistent ionospheric model, merely a convenient way to inject ions with a reasonable distribution with altitude than is dependent on solar zenith angle. While the scale height is used to inject particles we are not modeling the ionospheric processes themselves, justifying the comparably large resolution. We discuss the impact of this choice in Section \ref{sec:discussion}.

Each simulation is run on a $240^3$ grid, with boundaries $\pm 4$ RM in X, Y, and Z leading to an overall resolution of $\Delta x = 113$ km. Each simulation was run for 60,000 time steps with $\Delta t = 0.005$, or $\sim 3$ solar wind crossing times.  The final results analyzed for this paper were averaged over several timesteps once the simulation reached steady state, in order to improve statistics in low particle density regions.

The coordinate system used to present results throughout this paper is Planet Stellar Electric (PSE) coordinate system. This planet-centered system is used so that despite varying the direction of the IMF, the corresponding direction of the motional electric field ($E_{SW} = -u \times B$) remains constant.The PSE coordinates are then defined such that $-\hat x$ is defined to be in the direction of the solar wind, $\hat z$ is the direction of the convection electric field, and $\hat y$ is the completion of a right handed coordinate system. The simulation runs were performed in the same coordinate system as in our earlier study \citep{doi:10.1029/2017JA025068}, which is similar to the PSO (Planet Stellar Orbital) system. Transformation from PSO to PSE is a rotation around the X axis.

Much of the subsequent analysis was performed using the volumetric analysis package yt \citep{2011ApJS..192....9T} and visualization system Paraview \citep{Ayachit:2015:PGP:2789330}.

\section{Results}
  \label{sec:results}

  \subsection{Magnetic Field Morphology}
\label{sec:mag_morph}

\begin{figure*}
	\includegraphics[width=\textwidth]{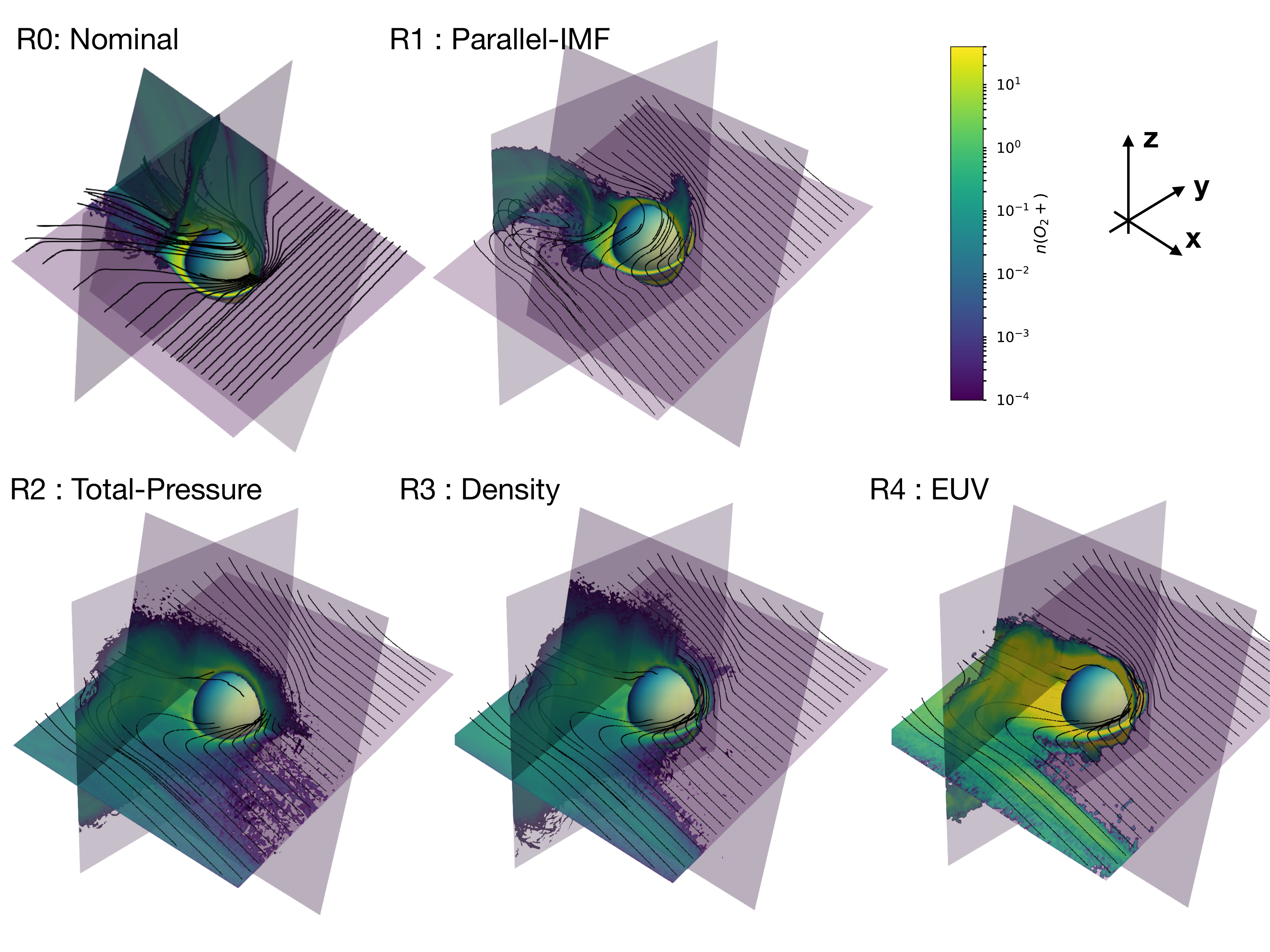}
	\centering
	\caption{Each panel shows slices of ${O_2}^+$ number density in the Z= 0, X=-1, and Y=0 planes with magnetic field lines traced in white. Panels show simulation R0, R1, R2, R3, and R4 from left to right, top to bottom. Note that the planes are not exactly aligned in each simulation due to the changing angle between induced electric field and magnetic field.}
	\label{fig:viz}
\end{figure*}

As discussed in Section \ref{sec:stellar_params}, the planetary interaction with the IMF and the resulting magnetic field configuration has key implications for ion escape. This intuition is confirmed in Figure \ref{fig:viz} which shows magnetic field lines traced through each simulation domain with slices of ${O_2}^+$  number density. 

The top two simulations in Figure \ref{fig:viz} show the difference between the quasi-perpendicular (left) and quasi-parallel (right) IMF. While the magnetic field lines pile up symmetrically in the quasi-perpendicular run, the pile-up is only significant in the +y hemisphere in the quasi-parallel run. In the -y hemisphere the magnetic field is much more bent close to the planet, leading to an offset s-shaped current sheet behind the planet. 

The location of the quasi-parallel shock determines where the magnetic field lines slip past the planet from their draped configuration to the current sheet. This location also corresponds to region where there is the greatest local curvature in the magnetic field close to the planet. While this region is symmetrically oriented over the +z pole in the R0 model, in the R1 model the region is offset towards the unstable shock side with the s-type current sheet.

Comparing simulations R1 and R2, the s-type current sheet becomes more extreme and the bend closer to the center of the planet gets much sharper. Additionally, some field lines that were clearly draped around the magnetic barrier in R1 \replaced{}{} appear to connect deeper in the magnetic barrier near the inner boundary (ionospheric obstacle) in R2. These differences are due to the much stronger magnetic field in the latter simulation. Because the magnetic field pressure is much stronger in the latter simulation it can more easily overcome the plasma pressure at low altitudes, embedding further field lines into the inner boundary. These field lines are thus less able to slip past the planet, extending the extent of the S-type current sheet. \added{As the magnetic field lines are pushed much deeper in the ionospheric region this may make our results sensitive to the conditions at the lower boundary.}

Simulations R3 and R4 show similar magnetic field morphologies to R2, despite the increases in solar wind density and ionospheric production rates, respectively.

  \subsection{Ion Morphology}
\label{sec:ion_morph}

\begin{figure*}
	\includegraphics[width=\textwidth]{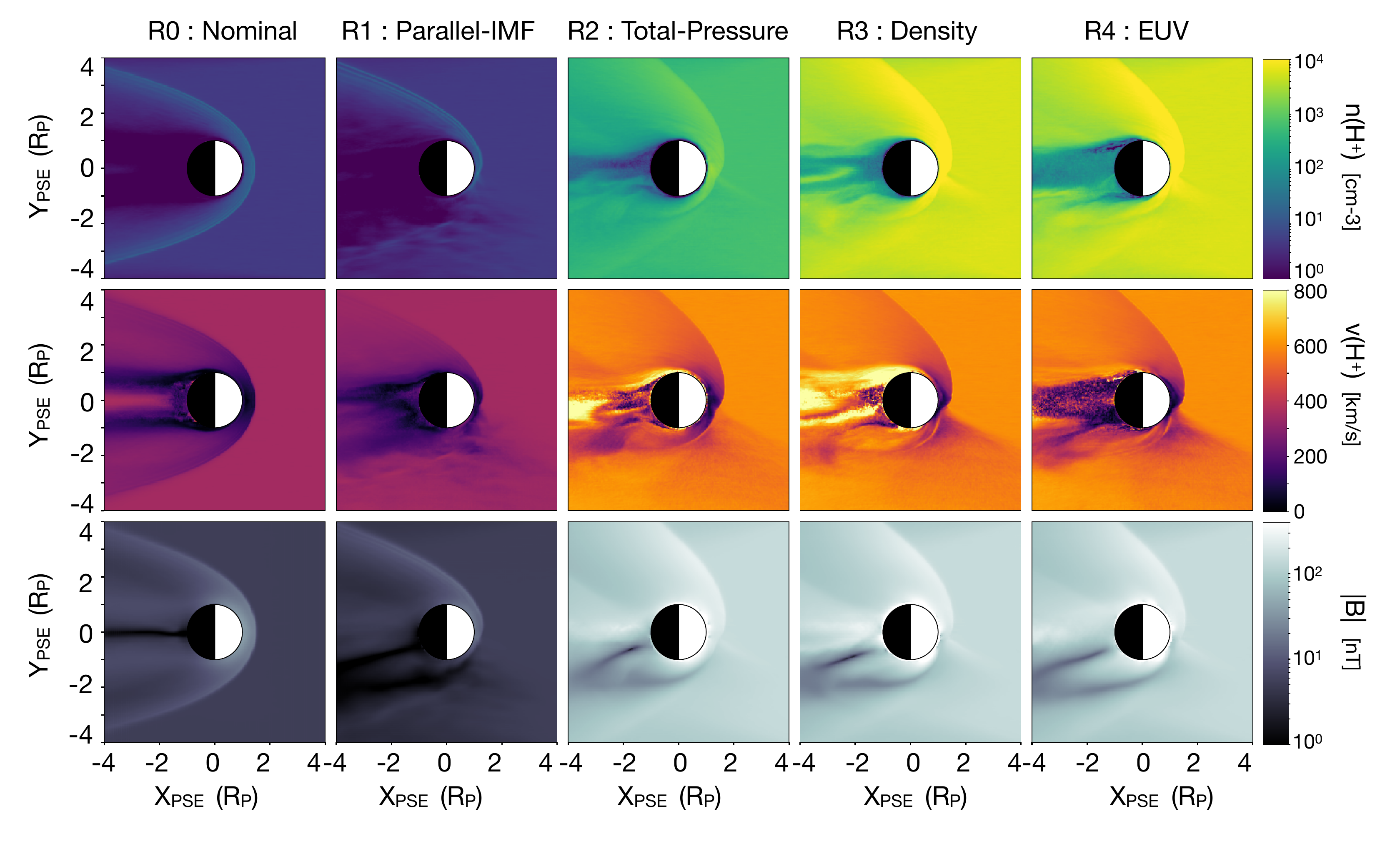}
	\centering
	\caption{Slices through the simulation domain at Z=0, showing the impact of the quasi-parallel shock. Here, the motional electric field is pointed out of the plane and the solar wind flows from right to left. Panels show H+ number density (top), H+ velocity (middle), and magnetic field magnitude (bottom) with identical color scales across all panels. From left to right the columns show simulations R0, R1, R2, R3, and R4.}
	\label{fig:qpar}
\end{figure*}

While overall ion loss rates are important for atmospheric evolution, this loss occurs through a variety of different processes, and it is important to understand the variation in each channel. Here we examine the overall ion morphology, and draw parallels to the ion escape channels seen at solar system objects and the different forces that govern the particle motion.

As discussed in section \ref{sec:mag_morph}, changing the IMF orientation from quasi-perpendicular to quasi-parallel drives asymmetry in the solar wind access near the ionosphere. In addition to creating an S-type current sheet, this asymmetry makes one hemisphere of the bow shock unstable as shown in Figure \ref{fig:qpar}, where slices of $H^+$ number density, $H^+$ velocity magnitude, and magnetic field magnitude are shown for each simulation. In each slice the solar wind flows from right to left, with the solar wind motional electric field normal to the plane. 

The unstable bow shock is evident in each row; the upper half of the bow shock shows sharply delineated boundary, while the lower half is ill-defined. This allows solar wind approaches the planet at a much higher velocity in lower hemisphere. This not only drives more energy transfer to the ionosphere, but drives ion pickup due to the $v \times B$ force from this location. 

\begin{figure*}
	\includegraphics[width=\textwidth]{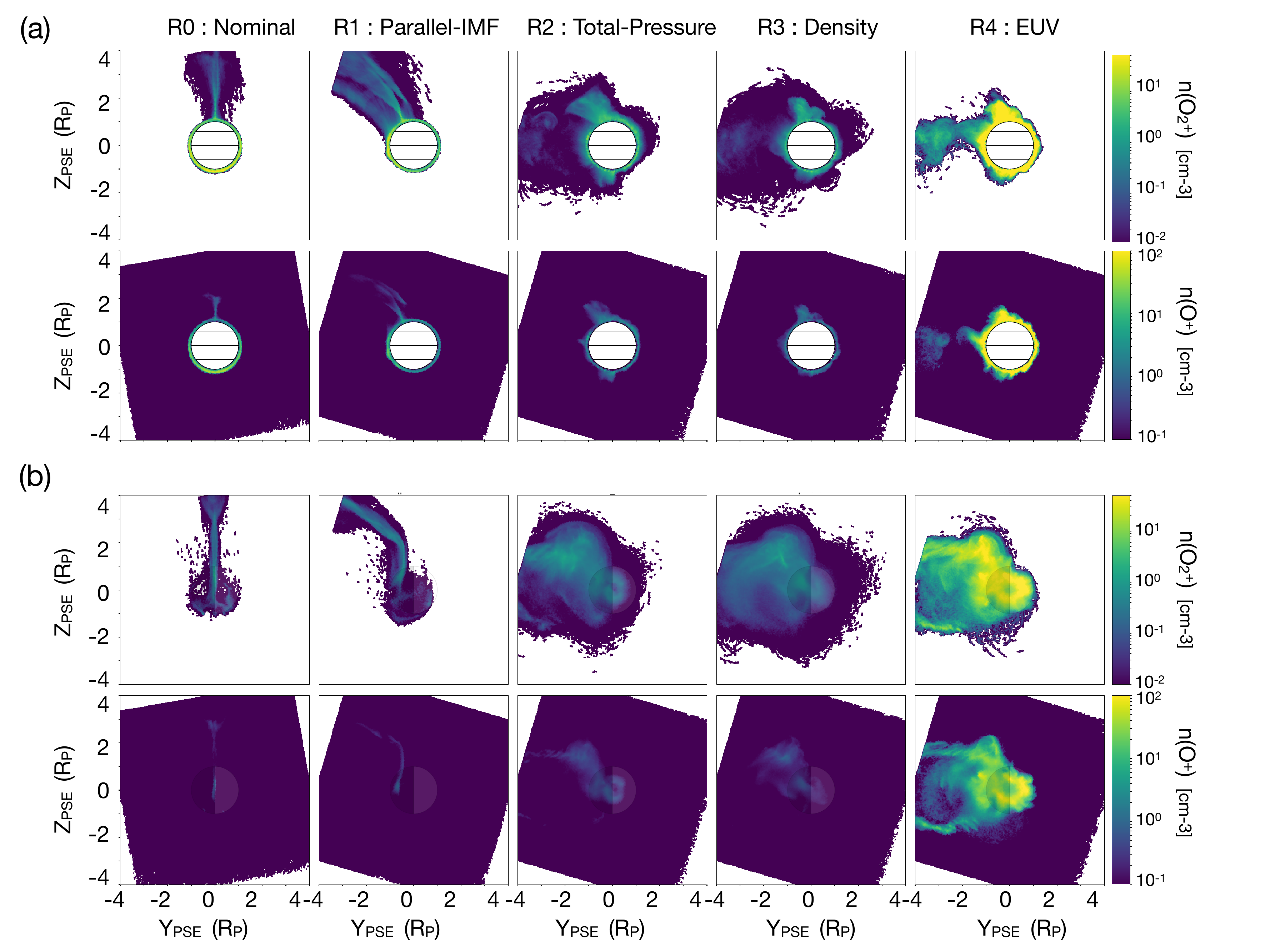}
	\centering
	\caption{Slices through the simulation domain at X=0 (a) and X=-1.5, showing heavy ion escape. Here, the motional electric field is pointed up and the solar wind flows normal to the page. Panels show O2+ number density (top), and O+ number density (bottom) with identical color scales across all panels. From left to right the columns show simulations R0, R1, R2, R3, and R4. The tilted box effects occur due to rotating the simulation domain into the PSE coordinate system.}
	\label{fig:plume}
\end{figure*}

Figure \ref{fig:plume} shows slices of the ${O_2}^+$ and ${O}^+$ number density for each simulation in the YZ plane. Comparing the first two panels illustrates the effect of the s-shaped configuration of the induced magnetosphere; while R0 shows symmetric acceleration in the direction of the motional electric field, R1 shows ions accelerated preferentially from the unstable shock hemisphere. While the ions in simulation R0 maintain their trajectory in the $+z$ direction along the symmetric current sheet, the ions in simulation R1 are redirected towards the asymmetric current sheet in the $-y$ hemisphere by the $J \times B$ force.

The morphology of escaping ions looks substantially less organized in the transition from R1 to R2. While the initial acceleration locations are the same, the outflow is much less collimated to the specific current sheet channel. This is due to much smaller gyroradii and changes in the current sheet configuration. As seen in Table 2, the large increase in the solar wind magnetic field with a modest increase in solar wind velocity drastically shortens the ion gyroradius to be much smaller than the size of the planet. Thus coherent motion on the scale of the planet is unlikely and the motion of even heavy ions like ${O_2}^+$ show magnetized behaviour. Furthermore, the changes in the current sheet discussed in section \ref{sec:mag_morph} have expanded the area from which ions are initially accelerated, broadening the eventual escape distribution. There is also a population of heavy ions that move upstream, after being quickly accelerated  from low latitudes on the day side. This is a relatively small population of particles as denoted by the low number densities, and do not contribute much to the overall escape.

Despite roughly an order of magnitude increase in solar wind number density from R2 to R3, the ion escape morphology remains roughly the same. This is likely a consequence of the similar magnetic field morphology. Similarly, when increasing the ion production rate by two orders of magnitude from R3 to R4, although the overall ion escape rates differ, the morphology of the escape again remains constant. 

  \subsection{Ion Escape}
\label{sec:ion_rates}

Table \ref{tab:ion_rates} lists a variety of metrics relating to ion escape rates for each simulation for both ${O_2}^+$ and $O^+$. Each of the escape properties were calculated by considering integrating the normal ion flux or power over a sphere located at $3.5 R_P$. This radius was chosen such that it is far enough from the planet that all ions are escaping and do not return back to the planet while not being affected by the simulation boundary. These results are roughly constant over $\pm 1 R_P$. The inflow power was calculated by integrating over the entire $+x$ simulation face. We chose to use the entire simulation face because the size of the bow shock approaches the size of the simulation domain.

Here we concentrate on the relative differences between the models, rather than the absolute magnitudes. Although this model has been validated by observations in solar system contexts \citep{2009AnGeo..27.4333J, RHybrid}, the specific escape rates are heavily dependent on the lower boundary conditions. As we are considering a generic exoplanet around an M-dwarf and there on not observed atmospheric constraints for any terrestrial exoplanets, we focus instead on the relative effects of the stellar wind conditions.

Each stage of stellar property changed increases the net escape flux of both ${O_2}^+$ and $O^+$, except for R2 to R3 (increasing the stellar wind dynamic pressure). The transition from quasi-perpendicular to quasi-parallel increases the amount of solar wind that can penetrate directly into the ionosphere. Increasing the solar wind strength in R1 to R2 increases the amount of energy that is put into the system, and the strength of the magnetic field used in the $v \times B$ and $J \times B$ forces to accelerate the ions. 

While the transition from R2 to R3 increases the solar wind dynamic pressure, and thus the total amount of energy available to the system, this does not translate to increased escape flux. Figure \ref{fig:qpar} shows that the increased density allows the solar wind to penetrate the ionosphere at higher velocity in some regions, which leads to an increased $v \times B$ force. However, the increased force does not lead to overall increased escape, because the escape is now production/diffusion limited. 

The escape fraction column of table \ref{tab:ion_rates} lists the fraction of total ions injected that escape the planet. While R0 and R1 have escape fractions of a few percent, R2 and R3 show that roughly $50\%$ of all injected ions are escaping. This limits the effect of the increased ion pickup force. The limit on ion escape is no longer the energy injected into the system, but the number of ions that are available to escape.

This is further illustrated in the transition to R4 when the ion production rate is increased.  While the overall escape flux increases, the escape fraction decreases as the production/diffusion limit is loosened. 

The next columns in Table \ref{tab:ion_rates} list the escape power, total inflow power, and the coupling constant $k$, defined as the ratio of the escape power to the inflow power.  The escape power follows roughly the same trends as the escape flux, except when comparing R0 and R1. In this case while the escape flux increases for a quasi-parallel IMF, the escape power slightly decreases. This is because the R1 heavy ions are not accelerated as much by the $v \times B$ force after leaving the planet, due to the much small perpendicular velocity component. 

The power coupling constant $k$ generally decreases as the stellar wind drivers are increased. The decrease from R0 to R1 corresponds to the decrease in escape power for the same stellar wind as discussed earlier. The power coupling also decreases from R1 to R2 as although the escape power increases, it does not increase as much as the solar wind power due to the limit of available ions. This effect is exacerbated in transition from R2 to R3, when the solar wind power continues to increase but the escape flux and power stay roughly constant. Finally, when the EUV is increased in R4 the power coupling constant increases, however, only to a rate comparable to R2, not as high as R0. Noting that the nominal case of R0 corresponds to the largest coupling constant is of key importance, because it implies that current observations of ion loss cannot be scaled indefinitely to more extreme conditions due to ion production/diffusion limitations. Thus, current observations may represent a more extreme case in solar wind power coupling.

The final columns in Table \ref{tab:ion_rates} list the average ion escape energy, or the escape power divided by the escape rate. For simulations R0 and R1 the energies are quite comparable to the average ion pickup energy expected given the solar wind parameters, as listed in Table \ref{tab:stellar_params}. For simulations R2-R4, however, the ions greatly exceed the estimates. This is because convection electric field is much stronger at lower altitudes than in the solar wind due to the increased dynamic pressure, leading to strong acceleration. The magnetic field draping allows there to be a larger component of the magnetic field perpendicular to the inflow velocity, plasma pressure balance ensures that the magnetic field pileup leads to strong magnetic fields, and higher bulk velocities are present closer to the planet as discussed in Section \ref{sec:ion_morph}. 

\begin{table*}
\begin{tabular}{@{}lllllllllll@{}}
\toprule
Simulation          & \multicolumn{2}{l}{Escape Flux}     & \multicolumn{2}{l}{Escape Fraction} & \multicolumn{2}{l}{Escape Power}  & Inbound Power & Power Coupling & \multicolumn{2}{l}{Escape Energy} \\
                    & \multicolumn{2}{l}{($10^{24}$ \#/s)} &                &                    & \multicolumn{2}{l}{($10^{10}$ W)} & ($10^{11}$ W) &                & \multicolumn{2}{l}{(keV)}         \\
                    & $O^+$          & ${O_2}^+$          & $O^+$          & ${O_2}^+$          & $O^+$          & ${O_2}^+$        &               &                & $O^+$         & ${O_2}^+$         \\\midrule
R0 : Nominal        & 0.6            & 0.6                & 0.04                    & 0.03                       & 0.2            & 0.3              & 2.5           & 0.02                            & 31            & 20.8              \\
R1 : Parallel-IMF   & 0.9            & 1.1                & 0.06                    & 0.05                       & 0.1            & 0.1              & 2.5           & 0.01                            & 5.6           & 6.9               \\
R2 : Total-Pressure & 6.4            & 9.6                & 0.45                    & 0.48                       & 4.5            & 1.4              & 1.7e3         & 0.001                           & 91            & 44                \\
R3 : Density        & 6.8            & 9.4                & 0.49                    & 0.47                       & 5.5            & 1.1              & 1.5e4         & 0.0001                          & 73            & 50                \\
R4 : EUV            & 400            & 410                & 0.29                    & 0.21                       & 190            & 404              & 1.5e4         & 0.003                           & 61            & 29                \\ \bottomrule
\end{tabular}
\caption{Ion escape flux, fraction, power, coupling to solar wind, and average escape energy for each simulation.}
\label{tab:ion_rates}
\end{table*}

\section{Discussion}
\label{sec:discussion}

\subsection{Model Limitations}
\label{sec:model_limitations}
\added{
One additional potential short-coming of the model we have applied here is that the ionospheric emission is driven by a predefined Chapman ion production profile. Accurately resolving ionospheric dynamics in the same domain as the global magnetosphere is computationally very challenging due to the large range of spatial and temporal scales; however, some simulation platforms include a one-way coupling from an ionosphere implementation to the global model, \citep[e.g.][]{2009JGRA..114.5216G,2016JGRA..12110190B,2016JGRA..121.6378M}.}

\added{We have also chosen to use a constant resistivity value above the lower boundary and zero resistivity at the lower boundary; however, a self-consistent model would couple the ionospheric electrodynamics and modulate the effective resistivity throughout the domain. Ionospheric resistivity is known to affect global thermosphere structure \citep{1982JGR....87.1599R} and ionospheric-magnetospheric coupling \citep{2004AnGeo..22..567R} through mechanisms such as current closure, atmospheric Joule heating, and Alfven wave dissipation. Furthermore, resistivity is dependent on auroral precipitation \citep{1987JGR....92.2565R, 1987JGR....92.7606F} and EUV flux \citep{1993GeoRL..20..971M}, both of which change across our simulations.}

\added{Modeling the ionospheric emission as a predefined production profile with a constant resistivity and the inner boundary as a super conducting sphere allows us to analyze stellar wind interactions of unmagnetized planets without an additional layer of uncertainty from a coupling between  ionospheric photo-chemistry, ionospheric electrodynamics, and global kinetic plasma models. These ionospheric models, while important, are poorly constrained with current upper atmospheric profiles of exoplanets. Furthermore, as the ion escape rates listed in Table 3 are not self-consistently resolved based on ionospheric photochemistry they should be taken as rough order of magnitude estimates. Further study should separately assess the variations of ionospheric production and electrodynamics with the change in stellar parameters considered here.}

\subsection{\added{Implications of} Changing Ion Loss Morphology}

In general, as the stellar input conditions are varied the morphology of the outflowing ions changes. The nominal case R1 showed symmetric tail and plume outflow from the nightside and mid-latitude dayside respectively. They were both collimated along the current sheet but well-defined as two different outflow channels. The R2 showed asymmetric outflow in both the tail and the plume due to the quasi-parallel shock and S-type current sheet. Models R3 and R4 showed outflow where the plume and tail were no longer distinct channels and were not well collimated. 

One immediate result from this is semantic; applying definitions of different ion outflow channels from solar system planetary science to exoplanets must be carefully considered. Although the initial acceleration mechanisms may be distinct, the outflow channels may not be.

Observable signatures may also vary as a result of different ion morphology. Although the possibility of observing such low density escape is far off, it is worth considering the wealth of different geometries that are possible.

Finally, different ion outflow morphologies may also have key implications for tidally locked planets. If heavy ions are preferentially accelerated from one hemisphere due to a quasi-parallel stellar wind interaction, rather than the day side of the planet, this may set up a diffusion limited scenario for escaping ions, or drive asymmetries in the environment at lower altitudes. 

\subsection{Ionospheric Loss \added{Rate Implications}}

Ion loss rates derived from simulations are often used to assess whether a planet is potentially habitable \citep[e.g.][]{2016arXiv160806919B}. While such rates may be validated by observations in solar system planetary contexts \citep{2009AnGeo..27.4333J,RHybrid}, the specific escape rates are heavily dependent on the ionospheric emission rates near the inner boundary, as discussed in Section \ref{sec:model_limitations}. Thus, beyond noting that the rates we find in our simulation cases R0-R3 are comparable to current ion loss rates derived for Earth, Venus, and Mars \citep{2005JGRA..110.3221S} and are thus relevant for discussing atmospheric evolution, we have focused our discussion on the relative difference in loss rates. 

\added{Atmospheric loss rates for the stellar parameters considered here may vary by several orders of magnitude; however, there is not a straightforward coupling between energy input and output, due to the complex coupling between the planet and the stellar wind. These results also imply that these systems are likely not energy limited. Instead, ion escape rates are likely limited by ion production or diffusion of the relevant species to the exobase.}

\deleted{
One additional potential short-coming of the model we have applied here is that the ionospheric emission is driven by a predefined Chapman ion production profile. Accurately resolving ionospheric dynamics in the same domain as the global magnetosphere is computationally very challenging due to the large range of spatial and temporal scales; however, some simulation platforms include a one-way coupling from an ionosphere implementation to the global model, \citep[e.g.][]{2009JGRA..114.5216G,2016JGRA..12110190B,2016JGRA..121.6378M}.}

\deleted{Modeling the ionospheric emission as a predefined production profile allows us to analyze stellar wind interactions of unmagnetized planets without an additional layer of uncertainty from a coupling between an ionospheric photochemistry model and a global kinetic plasma model and from poorly constrained upper atmospheric neutral profiles of exoplanets. The approach means that ion escape rates listed in Table 3 are not self-consistently resolved starting from intrinsic ionospheric processes and scales and should be taken as roughly order of magnitude estimates. In the analysis we have concentrated on plasma physics of exoplanetary induced magnetospheres and intercomparison of heavy ion escape between our simulation cases.}

\deleted{The hybrid model approach allows us to resolve 3-dimensional ion velocity distributions in the modeled induced magnetosphere down to the exobase. This information can be used when estimating, for example, the effect of the penetration of the solar wind ions along the S shaped open field lines near the upper atmosphere. The asymmetric solar wind ion precipitation may cause changes in the neutral atmosphere, which may have an impact in the ion escape.}

\section{Conclusions}
\label{sec:conclusions}

The plasma environment for potentially habitable planets around M-dwarfs is markedly different that the environment experienced by solar system planets like Venus, Earth, or Mars. Here we have presented a systematic study of the difference in environment and implications on magnetic field morphology and ion loss. The differences we considered were a quasi-parallel IMF orientation (R1), overall stellar wind pressure (R2), ratio of magnetic pressure to dynamic pressure in the solar wind (R3), and EUV input (R4). 

We found that both the ion loss morphology and overall loss rates were dictated by the plasma environment and magnetic field morphology. In cases where the stellar wind pressure was increased, the ion loss began to be diffusion- or production-limited with roughly half of all produced ions being lost. Because of this limit, the coupling of solar wind power to escaping ion power decreased in these extreme cases, despite the overall increase in ion loss. It is thus important to consider under what conditions scaling laws derived by observations of terrestrial planets begin to break down when applied to more extreme environments. 

Going forward, careful models of stellar winds for relevant systems will become increasingly important to constrain the plasma environment for potentially habitable exoplanets. Furthermore, it will be important to consider the dynamics of these systems, not only through an orbit of a steady state solar wind, but the intrinsic variability of any wind.

\section{Acknowledgments}
Hilary Egan was supported by the Department of Energy Computational Science Graduate Fellowship. This work utilized the Summit supercomputer and NERSC supercomputer. The Summit computer is supported by the National Science Foundation (awards ACI-1532235 and ACI-1532236), the University of Colorado Boulder, and Colorado State University. The Summit supercomputer is a joint effort of the University of Colorado Boulder and Colorado State University. This research used resources of the National Energy Research Scientific Computing Center (NERSC), a U.S. Department of Energy Office of Science User Facility operated under Contract No. DE-AC02-05CH11231. Global hybrid simulations were performed using the RHybrid simulation platform, which is available under an open source license by the Finnish Meteorological Institute (https://github.com/fmihpc/rhybrid).

\bibliographystyle{apj}
\bibliography{stellar_inf}

\end{document}